\documentclass[12pt]{iopart}
 
\usepackage{hyperref}
\usepackage{amsfonts}
\usepackage{amssymb}
\usepackage{subfig}
\usepackage{graphicx}
\usepackage{verbatim}

 \newcommand{\be}{\begin{equation}}
 \newcommand{\ee}{  \end{equation}}
  \newcommand{\ba}{\begin{array}}
  \newcommand{\ea}{\end{array}}
\newtheorem{theorem}{Theorem}
 \newtheorem{proposition}{Proposition}

\newcommand{\pD}[2]{\frac{\partial{#1}}{\partial{#2}}}

\newcommand{\cro}[1]{\left[ {#1} \right]}

\newcommand{\paren}[1]{\left( {#1} \right)}
\newcommand{\p}[1]{\left( {#1} \right)}
\newcommand{\de}[1]{\!\paren{#1}}

\newcommand{\prun}{\mathrm{pr}\,}

\newcommand{\al}{\alpha}

\newcommand{\la}{\lambda}
\newcommand{\La}{\Lambda}

\newcommand{\si}{\sigma}

\def\R{\mathbb{R}}

\begin{document}

\title{Soliton surfaces and generalized symmetries of integrable systems}
\author{A.M. Grundland$^{1,2}$, S. Post $^{3}$ and D. Riglioni$^1$}

\address{$^1$Centre de Recherches Math\'ematiques, Universit\'e de Montr\'eal,\\  C.P. 6128, Succ. Centre-ville, Montr\'eal, Quebec, H3C 3J7, Canada}
\address{$^2$Department of Mathematics and Computer Science ,\\ Universit\'e du Qu\'ebec, Trois-Rivi\`eres CP 500, Quebec, G9A 5H7, Canada}
\address{$^3$ Department of Mathematics, University of Hawaii, Manoa 2565 McCarthy Mall, Honolulu HI 96822 USA}
\ead{ grundlan@crm.umontreal.ca, spost@hawaii.edu  and riglioni@crm.umontreal.ca}
 
\submitto{\JPA}

\begin{abstract}
In this paper, we discuss some specific features of symmetries of integrable systems which can be used to contruct the Fokas-Gel'fand formula for the immersion of 2D-soliton surfaces, associated with such systems, in Lie algebras. We establish a sufficient condition for the applicability of this formula. This condition requires the existence of two vector fields which generate a common symmetry of the initial system and its corresponding linear spectral problem. This means that these two fields have to be group-related and we determine an explicit form of this relation. It provides a criterion for the selection of symmetries suitable for use in the Fokas-Gel'fand formula. We include some examples illustrating its application. \end{abstract}

\section{Introduction}
For the last two centuries, Lie group theory, as a theory of transformations of sets of solutions of differential equations \cite{Lie S.}, has generated a great deal of interest and activity in several areas of mathematics and physics. Over those years, the development of this theory has led to significant progress in classifying  and solving differential equations, yielding a wealth of new interesting results. A number of attempts to generalize this subject and develop its applications can be found in the literature (see e.g. \cite{Cartan}-\cite{Mikhailov} and references therein). 
Of particular interest, from the physical point of view, has been development of the theory of integrable (or soliton) surfaces \cite{bob1990} arising from infinitesimal deformations of integrable differential equations and describing the behaviour of soliton solutions. Construction and analysis of these surfaces, associated with integrable systems in such areas as field theory, quantum mechanics and fluid dynamics have provided new tools for investigating nonlinear phenomena described by these systems (for review of this subject see e.g. \cite{rogers,babelon,manton}).

The possibility of using a linear spectral problem (LSP) to represent a moving frame on an integrable surface has yielded many new results concerning the intrinsic geometric properties of such surfaces (see e.g.  \cite{BobEitbook} - \cite{sym1985soliton}). In this regard, it has recently proved fruitful to extend such a characterization of soliton surfaces via their immersion functions in Lie algebras. The construction of such surfaces, related to integrable models, has been accomplished (see e.g. \cite{fokas1996surfaces}-\cite{baran} ) by representing the equation of motion for the model by a matrix zero-curvature representation (ZCR). The problem of the determination of an explicit expression for an immersion function of 2D-surfaces in Lie algebras associated with integrable systems of partial differential equations was first solved by A. Sym \cite{sym1985soliton} and J. Tafel \cite{tafel1995surfaces} in 1985 using a conformal symmetry in the spectral parameter $\lambda\in\mathbb{C}$ of the integrable system. Since then several generalizations of the immersion formula for the 2D-surfaces have been introduced, based on gauge symmetries by J. Cieslinski \cite{cieslinski1997ageneralized},\cite{cieslinski} and A. Doliwa \cite{doliwa1992constant} and, more recently, built on generalized infinitesimal symmetries of the integrable equations by A. Fokas and I. Gel'fand \cite{fokas1996surfaces}, and later with Finkel and Liu \cite{fokas2000aformula}. The latter one is the most general up to date, providing explicit formula for generating infinitely many surfaces in Lie algebras. This subject was further developed by several authors (see e.g. \cite{Ceyhan} - \cite{grundland2012soliton}), among them two of the authors of this paper (\cite{grundland2011soliton} - \cite{grundland2012soliton}). These studies resulted in the determination of new forms of bijective correspondence between various symmetries of the integrable system and 2D-surfaces immersed in a Lie algebra.

In this paper we continue our work on the application and analysis of the results obtained by Fokas et al. in \cite{fokas1996surfaces}, \cite{fokas2000aformula}. We formulate a sufficient condition for the existence and explicit integration of soliton surfaces in Lie algebras obtained through a common symmetry of both the integrable system and its corresponding LSP. This can be accomplished by the use of the generalized symmetry formalism based on vector fields written in evolutionary form, which are defined on some extended jet space, their prolongation structure and their connections with the formal Fr\'echet derivatives. In particular, the integrable equation and the LSP determine (possibly different) conditions on the jet space. Taking the quotient of the jet space with respect to these conditions gives two (possibly different) varieties. A generalized symmetry is a common symmetry of both the integrable equation and the LSP if and only if it can be represented by a tangent vector field to both varieties. We show that this is equivalent to the requirement that this vector field is group related (in the sense given in \cite{olver1993application}) to a vector field on the jet space defined by the LSP.

The plan of this paper is as follows. Section 2 contains a brief account of basic definitions and properties concerning the immersion of soliton surfaces associated with integrable equations and fixes the notation. In section 3, we present a geometric formulation of the Fokas-Gel'fand formula for the immersion of 2D-surfaces in Lie algebras. Next, we determine a sufficient condition for the existence of an integrated form of the surfaces obtained via a Lie algebra of common symmetries of an integrable system and its LSP. We establish an explicit requirement for the vector fields associated with symmetries of the ZCC and the LSP which ensures that they generate a common symmetry.
In sections 4 and 5 we present examples of applications of the proposed approach to the Jacobi elliptic equations and to the $CP^{N-1}$ sigma models, respectively. We investigate whether the Fokas-Gel'fand formula for immersion can or cannot be integrated explicitly. Section 6 contains final remarks.

\section{Preliminaries on generalized symmetries of integrable systems}
In this section we collect several facts concerning the immersion of 2D-surfaces associated with integrable systems in two independent variables and discuss the decomposition of this immersion into an appropriate Lie algebra. We conduct our analysis in terms of vector fields and their prolongations and we use the standard notation of prolongation formalism as proposed by P. J. Olver \cite{olver1993application}. In what follows, we have adopted the convention that repeated indices are summed unless one of them is in a bracket. For simplicity we assume that all considered functions, tensor fields and manifolds are of class $C^\infty$. All our considerations are of a local character. 

Let us consider a set of functions $\lbrace\theta^k\rbrace_1^q$ of two independent variables 
$\xi = (\xi_1, \xi_2) $. For the $n$-th order partial derivatives we use the multi-index notation 

\begin{equation}
\partial_{J} = \frac{\partial^n }{\partial \xi^{j_1} ... \partial  {\xi^{j_n}}} , 
\end{equation}
 
\noindent where $J = (j_1, ... , j_n)$ is an unordered $n$-tuple of integers with entries $j_i=1,2$ and $ \# J = n $. The derivatives of $\theta^k$ up to order $n$ are denoted by $\theta_{J}^k = \partial_J \theta^k (\xi) $, which represent all the different $n$-th order derivatives of $\theta^k$ at the point $\xi$. For simplicity we use the abbreviated notation of the jet space $N\ni[\theta]=(\xi_1,\xi_2,\theta^k,\theta^k_J)$, where $\xi_1,\xi_2,\theta^k,\theta^k_J$ play the role of coordinates of $N$. 

A generalized vector field written in evolutionary representation and its prolongation have the form  \cite{olver1993application} 

\begin{equation}
\omega_R=R^k[\theta]\frac{\partial}{\partial\theta^k},\qquad \textrm{pr}\omega_R=R^k[\theta]\frac{\partial}{\partial\theta^k}+\left(D_JR^k[\theta]\right)\frac{\partial}{\partial\theta^k_J},
\end{equation}
where $D_J$ is the total derivative operator defined iteratively by
\begin{equation}
D_{J, \alpha} = D_\alpha D_J, \quad D_\alpha =\frac{\partial}{\partial\xi_\alpha}+\theta^k_{J,\alpha}\frac{\partial}{\partial\theta^k_J} ,\qquad \alpha=1,2.\label{total}
\end{equation}

Let us denote by $G$ a finite-dimensional Lie group and by $\mathfrak{g}$ its corresponding Lie algebra $(\textrm{dim} \quad \!\!\!\!\! \mathfrak{g}  <\infty$). 
The bracket $[\cdot,\cdot]$ denotes the Lie algebra commutator, i.e.
\begin{equation}
[e_i,e_j]=c^k_{ij}e_k,
\end{equation}
where $c_{ij}^k$ are the structural constants of the Lie algebra $\mathfrak{g}$. In the language of the jet space $N$ a system of partial differential equations (PDEs) in two independent variables $\xi_1,\xi_2$ and dependent variables $\theta^k(\xi_1,\xi_2)$ is denoted by
\begin{equation}
\Omega[\theta]=0.\label{initial}
\end{equation}
Let us suppose that the system \eref{initial} can be linearized by a matrix linear spectral problem (LSP) given by
\begin{equation}
\Lambda([\theta],\lambda)\equiv D_\alpha\Phi([\theta],\lambda)-U^\alpha([\theta],\lambda)\Phi([\theta],\lambda)=0,\qquad \alpha=1,2\label{LSP}
\end{equation}
whose compatibility conditions are in the form of a zero-curvature condition (ZCC) which is assumed to be independent of the spectral parameter $\lambda\in\mathbb{C}$,
\begin{equation}
\Omega[\theta]\equiv D_2U^1-D_1U^2+[U^1,U^2]=0.\label{ZCC}
\end{equation}
We call such a representation of the integrable system of differential equations a zero-curvature representation. The potential matrices $U^\alpha([\theta],\lambda)$ take values in a Lie algebra $\mathfrak{g}$ and the differentiable function $\Phi([\theta],\lambda)$ takes values in its corresponding Lie group $G$.

In \cite{fokas2000aformula}, dealing with general forms of smooth 2D-surfaces immersed in the Lie algebra $\mathfrak{g}$, whose Gauss-Mainardi-Codazzi (GMC) equations are equivalent to an infinitesimal deformation of an integrable equation $\Omega[\theta]=0$, Fokas, Gel'fand, Finkel and Liu look for a simultaneous infinitesimal deformation of the LSP \eref{LSP} and of the ZCC \eref{ZCC}. This means that for any infinitesimal change of the form
\begin{equation}
\left[\ba{c}
(U^1)' \\
(U^2)'\\
\Phi'
\ea\right]=\left[\ba{c}
U^1\\
U^2\\
\Phi
\ea\right]+\epsilon\left[\ba{c}
A^1\\
A^2\\
\Psi=\Phi F
\ea\right],\qquad 0<\epsilon\in\mathbb{R},\label{defor}
\end{equation}

\noindent (where $A^1, A^2$, $\Psi$ and $F$ take their values in a Lie algebra $\mathfrak{g}$) the matrix functions $( (U^1)' , (U^2)' ,\Phi' )^T$have to satisfy the same singularity structure in the spectral parameter $\lambda$ as $(U^1,U^2,\Phi)^T$ . The infinitesimal deformation of the ZCC \eref{ZCC} requires that the matrix functions $A^1$ and $A^2$ satisfy
\begin{equation}
D_2A^1-D_1A^2+[A^1,U^2]+[U^1,A^2]=0,\label{deter}
\end{equation}
and the infinitesimal deformation of the LSP \eref{LSP} implies
\begin{equation}
D_1\Psi=U^1\Psi+A^1\Phi,\qquad D_2\Psi=U^2\Psi+A^2\Phi.\label{psi}
\end{equation}
 
\noindent The requirement (\ref{deter})  coincides with the compatibility condition for (\ref{psi}). Hence the existence condition for a $\mathfrak{g}$-valued immersion function $F=\Phi^{-1}\Psi$ of a 2D-surface in a Lie algebra $\mathfrak{g}$ can be expressed in terms of the matrix functions $A^1$ and $A^2$ which satisfy (\ref{deter}) and (\ref{psi}). This result can be formulated as follows 

\begin{theorem}[A. Fokas et al. \cite{fokas1996surfaces}, \cite{fokas2000aformula},  our formulation] Suppose that there exist two $\mathfrak{g}$-valued matrix functions $U^1$, $U^2$ and a $G$-valued function $\Phi$ chosen so as to satisfy the ZCC \eref{ZCC} and its associated LSP \eref{LSP}. Suppose also that the $\mathfrak{g}$-valued matrix functions $A^1$ and $A^2$ are linearly independent and satisfy \eref{deter}. Then, there exists a 2D-surface with $\mathfrak{g}$-valued immersion function $F([\theta],\lambda)$ such that the tangent vectors are given by
\begin{equation}\label{tan}
D_\alpha F([\theta],\lambda)=\Phi^{-1}A^{\alpha}([\theta],\lambda)\Phi,\qquad\alpha=1,2.
\end{equation}
The first and second fundamental forms of the surface are expressible in terms of the matrices $U^\alpha$ and $A^\alpha$ only.
\end{theorem}

Equation (\ref{deter}) is an infinitesimal deformation of the ZCC  $\Omega [\theta] = 0$ and such an infinitesimal deformation can be generalized by three choices of symmetries, namely generalized symmetries of the ZCC  $\Omega [\theta] = 0$ considered as a PDE in the matrix variables $U^1$ and $U^2$, generalized symmetries of the differential equations $\Omega [\theta] = 0$ in unknown functions $\theta$'s and symmetries with respect to the spectral parameter $\lambda$. The surfaces generated by a linear combination of such symmetries are called surfaces associated with integrable systems.

In their study, Fokas et al. \cite{fokas2000aformula} consider admissible classes of symmetries of the PDE \eref{initial}, including a conformal translation in the spectral parameter $\lambda\in\mathbb{C}$, a gauge transformation of the function $\Phi\in G$ and a set of generalized symmetries of the integrable system \eref{initial}. Based on these symmetries they show that the $\mathfrak{g}$-valued immersion function $F$ of a 2D-surface can be constructed explicitly through the integration of tangent vectors \eref{tan} when $\mathfrak{g}$-valued matrix functions $A^\alpha$ have the form
\begin{equation}
\fl A^\alpha([\theta],\lambda)=\beta(\lambda)D_\lambda U^\alpha+D_\alpha S+[S,U^\alpha]+\frac{DU^\alpha}{D\theta^j}R^j\in\mathfrak{g},\qquad\alpha=1,2,\label{A}
\end{equation}
where $({DU^\alpha}/{D\theta^j})R^j$ is the Fr\'echet derivative of $U^\alpha$ in the direction of the scalar functions $\lbrace R^j\rbrace_1^n$ which are symmetries of the equation \eref{initial}. Here $\beta(\lambda)$ is an arbitrary function of a spectral parameter $\lambda$ and $S([\theta],\lambda)$ is a $\mathfrak{g}$-valued matrix function defined on the jet space $N$ and depending on the spectral parameter $\lambda$. Fokas et al. propose a correspondence between symmetries of an integrable equation \eref{initial} and 2D-surfaces immersed in the Lie algebra $\mathfrak{g}$ which is governed by the following formula (up to an additive $\mathfrak{g}$-valued constant)
\begin{equation}
F([\theta],\lambda)=\Phi^{-1}\left(\beta D_\lambda\Phi+S\Phi+\frac{D\Phi}{D\theta^j}R^j\right)\label{Phi}.
\end{equation}
The integrated form \eref{Phi} defines a mapping $F:\mathbb{R}^2\rightarrow\mathfrak{g}$. In what follows, we will refer to it as the Fokas-Gel'fand (FG) formula for immersion. The three terms correspond to conformal transformations of the spectral parameter (the Sym-Tafel formula for immersion), gauge symmetry of the LSP, and generalized symmetries of the PDEs. Here, and for the remainder of this paper, we retain the symbol $F$ for the immersion function reduced to its third term, i.e. for the case when $\beta$ and $S$ in \eref{Phi} are equal to zero, i.e. 
\begin{equation}
F=\Phi^{-1}\frac{D \Phi }{D \theta^{j}} R^j.\label{f}
\end{equation}
In this case matrices $A^\alpha$ become
\begin{equation}
A^{\alpha}([\theta ],\lambda)=\frac{DU^{\alpha}}{D\theta^j}R^j.\label{a}
\end{equation}

\section{The sufficient conditions}
\label{theory}

In this paper we continue the line of inquiry developed in \cite{grundland2011soliton}, where we established some limitations on the applicability of the formula \eref{Phi}. We proved that the
expression for a surface given by (\ref{f}), together with tangent vectors (\ref{tan}), holds if and only if there exists a common symmetry of the ZCC $\Omega[\theta]=0$ and its corresponding LSP $\Lambda([\theta],\lambda)=0$. However, applying this criterion is challenging from the technical point of view (since it requires a verification of whether the action of $\textrm{pr}\omega_R$ on the LSP vanishes for all solutions of the LSP). That has motivated us to re-examine in depth the problem of the sufficiency condition for the existence of an immersion function $F$ in the form \eref{f}. Our investigation, presented in this paper, has led us to a reformulation of this condition in the language of vector fields and their prolongations which in effect has provided us with a more efficient way of verifying the existence of common symmetries of the ZCC $\Omega[\theta]=0$ and its LSP $\Lambda([\theta],\lambda)=0$.

To proceed with our analysis let us first recapitulate the main result of \cite{grundland2011soliton} for which we present a more direct proof, better suited to our purpose here.

As demonstrated in \cite{grundland2011soliton}, the validity of the formula \eref{Phi} depends crucially on its third term involving generalized symmetries of the initial equation \eref{initial}. A surface with $\mathfrak{g}$-valued function $F$ given by \eref{f} exists if and only if the generalized symmetry  $\omega_R$ is a common symmetry of both the system ZCC $\Omega[\theta]=0$ and its corresponding LSP $\Lambda([\theta],\lambda)=0$. To demonstrate this fact we make use of the Lie group structure of the generalized symmetries, their related vector fields, and their links to total and Fr\'echet derivatives. We use the prolongation formalism of the vector fields instead of Fr\'echet derivatives, as it provides an effective computational procedure for finding the most general symmetry group of any system of PDEs under consideration \cite{olver1993application}. A vector field written in the evolutionary form $\omega_R$ defined on jet space $N$ 
\begin{equation}
\omega_R=R^k[\theta]\frac{\partial}{\partial\theta^k}\label{omega}
\end{equation}
is a symmetry of the system ZCC $\Omega[\theta]=0$ if and only if the first prolongation of this system vanishes, i.e. 
\begin{eqnarray}
\textrm{pr}\omega_R\left(D_2U^1-D_1U^2+[U^1,U^2]\right)\label{pr} \\  = D_2 (\textrm{pr}\omega_R U^1) - D_1 (\textrm{pr}\omega_R U^2) + [\textrm{pr}\omega_R U^1 , U^2] + [U^1, \textrm{pr}\omega_R U^2] = 0, \nonumber
\end{eqnarray}
whenever \eref{ZCC} holds. The expression \eref{pr} is equivalent to \eref{deter} with
\begin{equation}
A^\alpha=\textrm{pr}\omega_RU^\alpha,\qquad\alpha=1,2\label{Aalpha}
\end{equation}
since the total derivatives $D_\alpha$ commute with the prolongation of the vector field $\omega_R$ in evolutionary form (\cite{olver1993application}, Lemma 5.12, p. 300), i.e.
\begin{equation}
[D_\alpha,\textrm{pr}\omega_R]=0,\qquad \alpha=1,2.\label{com}
\end{equation}

The Fr\'{e}chet derivative of $\Phi$ with respect to $\theta^k$ can be expressed through the prolongation of the vector field $\omega_R$ (\cite{olver1993application}, Proposition 5.25 in  p.307). Thus, we can obtain \cite{grundland2011soliton} the following equivalence for the formula \eref{f}
\begin{equation}
F=\Phi^{-1}\frac{D\Phi}{D\theta^k}R^k=\Phi^{-1}(\textrm{pr}\omega_R\Phi).\label{eq19*}
\end{equation}

\noindent Differentiating  \eref{f} and using the LSP \eref{LSP}, we obtain
\begin{eqnarray}
\mathrm{D}_{\al}F=\mathrm{D}_{\al}\p{\Phi^{-1}\frac{\mathrm{D}\Phi}{\mathrm{D}\theta^n}R^n}=\Phi^{-1}\cro{-U^{\al}\frac{\mathrm{D}\Phi}{\mathrm{D}\theta^n}R^n+\mathrm{D}_{\al}\p{\frac{\mathrm{D}\Phi}{\mathrm{D}\theta^n}R^n}}.\label{eq9}
\end{eqnarray}
Making use of the relations \eref{com} and \eref{eq19*}, we can write the second term in \eref{eq9} as
\begin{eqnarray}
\mathrm{D}_{\al}\p{\frac{\mathrm{D}\Phi}{\mathrm{D}\theta^n}R^n}=\mathrm{D}_{\al}\p{\prun \omega_R\Phi}=\prun \omega_R\p{\mathrm{D}_{\al}\Phi}.\label{eq10}
\end{eqnarray}
Using the identity
\begin{eqnarray}
\prun \omega_R\p{\mathrm{D}_{\al}\Phi}=\prun \omega_R\p{U^{\al}\Phi} + ( \prun \omega_R\p{\mathrm{D}_{\al}\Phi-U^{\al}\Phi} ) ,\label{eq12}
\end{eqnarray}
we determine that the second term in the right hand side of \eref{eq12} is not necessarily zero. This term vanishes if and only if the vector field $\omega_R$ is also a symmetry of the LSP $\La\de{\cro{\theta},\la}=0$ in the sense that
\begin{equation}
\textrm{pr}\omega_R(D_\alpha\Phi-U^\alpha\Phi)=0,\qquad\textrm{whenever}\qquad D_\alpha\Phi-U^\alpha\Phi=0\label{eq20*}.
\end{equation}
Let us assume that \eref{eq20*} holds. Then from \eref{eq12} we obtain
\begin{eqnarray}
\prun \omega_R\p{\mathrm{D}_{\al}\Phi} = \prun \omega_R\p{U^{\al}\Phi}&=\p{\prun \omega_R U^{\al}}\Phi+U^{\al}\p{\prun \omega_R\Phi}\nonumber\\
&=\p{\frac{\mathrm{D} U^{\al}}{\mathrm{D}\theta^n}R^n}\Phi+U^{\al}\p{\frac{\mathrm{D}\Phi}{\mathrm{D}\theta^n}R^n}.\label{eq13}
\end{eqnarray}
Substituting \eref{eq13} into \eref{eq9} and using \eref{Aalpha} we obtain the tangent vectors $\mathrm{D}_{\al}F$ in the form postulated by Theorem 1
\begin{eqnarray}
\mathrm{D}_{{\al}}F&=\Phi^{-1} \cro{-U^{\al} \p{\frac{\mathrm{D}\Phi}{\mathrm{D}\theta^n}R^n} +\p{\frac{\mathrm{D}U^{\al}}{\mathrm{D}\theta^n}R^n}\Phi +U^{\al}\p{\frac{\mathrm{D}\Phi}{\mathrm{D}\theta^n} R^n} }\\
&=\Phi^{-1} \p{ \frac{\mathrm{D}U^{\al}}{\mathrm{D}\theta^n}R^n}\Phi = \Phi^{-1} (\mathrm{pr} \omega_R U^{\alpha})\Phi=\Phi^{-1}A^{\alpha}([\theta],\lambda)\Phi\nonumber. \label{eq14}
\end{eqnarray}
Thus, under the condition that the vector field $\omega_R$ is also a symmetry of the LSP, there exists a 2D-surface with $\mathfrak{g}$-valued immersion function given by \eref{eq19*}. For the remainder of this section, we investigate the meaning of this condition.

Let us recall that the $\mathfrak{g}$-valued immersion function $F$ is defined via an infinitesimal deformation (\ref{defor}) of the LSP (\ref{LSP}) and that the ZCC (\ref{ZCC}) is treated as a differential equation for $\Phi$ and $U^\al$. In terms of the formalism of vector fields on a jet space, we have the jet space $M=[\Phi, U]$ which is defined locally by the components of $\Phi$ and $U^\al$. That is, the components $U^{\al}$ are of the form $U^{\al,j}e_j$ and, for the function $\Phi$ which takes values in a neighbourhood $\mathcal{B}$ of the unit element $e\in G$, we use the canonical coordinates defined by the formula

\begin{eqnarray}
\Phi =\exp (\Phi^j e_j)\in G.\label{eq20}
\end{eqnarray}
For other points, which do not belong to $\mathcal{B}$, one constructs a similar parametrization in some neighbourhood of a point $\Phi_0$. Using left translation, we have
\begin{eqnarray}
\Phi=\exp (\Phi^je_j)\Phi_0.\label{eq21}
\end{eqnarray}
The vector field in evolutionary form defined on the jet space M takes the form 
\be \eta_Q=Q^j[\Phi, U]\frac{\partial}{\partial \Phi^j}+\hat{Q}^{\al ,j}[\Phi, U]\frac{\partial }{\partial U^{\al ,j}}.\ee
The vector field $\eta_Q$ is assumed to be a generalized symmetry of the LSP in the sense that 
\be pr \eta_Q\left(D_\al \Phi-U^\al \Phi\right)=0, \qquad \mbox{ whenever } D_\al \Phi-U^\al \Phi=0, \quad \al = 1,2 . \ee
Here, we are considering the LSP as a set of conditions on the jet space M, namely the  conditions 
\be \Sigma^{\al,j}[\Phi, U]=\left[ (D_\al \Phi)\Phi^{-1}-U^\al \right]^j=0,\ee
where the notation $[\cdot ]^j$ denotes the component of the argument in the j-th basis element of the algebra $\mathfrak{g}$. In terms of the infinitesimal symmetry (\ref{defor}), we have 
\begin{eqnarray} 
A^1=pr\eta_Q U^1, \qquad 
A^2=pr\eta_Q U^2 \label{Ai}, \qquad 
\Psi=pr\eta_Q \Phi=\Phi F.
\end{eqnarray}
From equations (\ref{Ai}), any symmetry of $\Sigma[\Phi, U]=0$ (i.e. when $A^1,A^2,\Psi$ satisfy equations (\ref{deter}) and (\ref{psi})) provides an immersed surface in the Lie algebra $\mathfrak{g}$ and, furthermore, the existence of this surface gives exactly the condition for such an infinitesimal deformation to exist. 

Now, we are investigating the case where the LSP is parametrized by some integrable equation and spectral parameter $\lambda$; that is, both $\Phi$ and $U^{\al}$ are functions on the jet space $[\theta]$ parametrized by $\lambda$. In this case, the LSP is given by the equation $\Lambda([\theta], \lambda)=0$ as above. Note that for a particular solution $\Phi([\theta], \lambda)$ of the LSP (\ref{LSP}), this equation reduces to a (possibly $\lambda$-dependent) differential equation for $\theta$ which is satisfied for all solutions of the considered integrable equation $\Omega[\theta]=0$. However, as is often the case, it may not be possible to give $\Phi$ in a closed form for arbitrary solutions. Or, it may be that even for arbitrary solutions $\Phi$, the equation $\Lambda([\theta], \lambda)=0$ does not have the same symmetry group as $\Omega[\theta]=0$. 

Geometrically, suppose that we have a solution $\Phi([\theta],\lambda)$ which satisfies $\Lambda([\theta], \lambda)=0$ whenever $\Omega[\theta]=0$. Then, we define two subvarieties of the jet space N. We denote by ${\cal O}$ the variety which consists of the points in $N$ subject to the constraint $\Omega[\theta]=0$, and by ${\cal L}$ the variety which consists of the points in $N$ subject to the constraint $\Lambda([\theta], \lambda)=0.$ A vector field $\omega_R$ is a symmetry of $\Omega[\theta]=0$ if and only if its prolongation lies in the tangent space of ${\cal O}$. It is then a simultaneous symmetry if and only if it also lies in the tangent space of ${\cal L}$. 

Furthermore, if we would like to relate the vector fields $\omega_R$ and $\eta_Q$, this exactly corresponds to the requirement that the vector fields be $\tau$-related  (in the sense introduced in \cite{olver1993application}, page 33). That is, for any parametrization of $\Phi([\theta], \lambda)$ and $U^\al ([\theta], \lambda)$ in terms of the jet space $N$ and a spectral parameter $\lambda$, we obtain a mapping from the jet space $N$ to the jet space $M$, which we call $\tau$. It is an admissible parametrization if and only if 
\be \mathrm{pr}\tau {(\omega_R)} \left(D_\al \Phi-U^{\al}\Phi\right)=0 , \quad \mbox{ whenever } \Omega[\theta]=0.\ee
The requirement that $\omega_R$ be $\tau$-related to $\eta_Q$ means exactly that the push-forward of $\omega_R$ should be the vector field $\eta_Q $ on the image of $\tau$, i.e.
\be d\tau(\omega_R)=\eta_Q\bigg|_\tau.\label{taurelated}\ee
In coordinates, this becomes 
\be d\tau(\omega_R)\equiv
\si_R = R^k\p{\cro{\theta}}\left( \pD{\Phi^j}{\theta^k}\pD{}{\Phi^j}+\pD{U^{\al ,j}}{\theta^k}\pD{}{U^{\al ,j}}\right) \bigg\vert_{\tau}\label{eq18}.
\ee
Thus the requirement that the vector fields be $\tau$-related (\ref{taurelated}) is equivalent to 
\begin{eqnarray}
Q^j|_\tau= R^k \cro{\theta}  \frac{\partial\Phi^j(\cro{\theta}, \lambda)}{\partial\theta^k}, \qquad \hat{Q}^{\al,j}\bigg|_\tau=R^k \cro{\theta} \pD{U^{\al,j}}{\theta^k}.\label{tau}
\end{eqnarray}

\noindent We are ready then to state the following. 
\begin{proposition} Suppose that there exist two vector fields, $\eta_Q$ on the jet space $M=[\Phi , U]$, which is a generalized symmetry of the general form of the LSP $\Sigma[\Phi, U]=0$, and $\omega_R$ on the jet space $N=[\theta]$, which is a generalized symmetry of the integrable PDE $\Omega[\theta]=0$. If these vector fields are $\tau$-related, then the vector field $\omega_R$ will be a common symmetry of both $\Omega[\theta]=0$ and  $\La\de{\cro{\theta},\la}=0.$
\end{proposition}
The proof of this theorem follows directly from the chain rule and the fact that $\eta_Q|_\tau=\sigma_R$. Namely,
\begin{eqnarray}
\mathrm{pr}{\omega}_R\left(D_\al \Phi-U^{\al}\Phi\right)&=&\mathrm{pr}\sigma_R\left(D_\al \Phi-U^{\al} \Phi\right)\\
&=&\mathrm{pr}\eta_Q\left(D_\al \Phi-U^{\al} \Phi\right)\bigg|_\tau \nonumber \\
&=&0, \qquad \mbox{ whenever } \Omega[\theta]=0\nonumber,
\end{eqnarray} 
which completes the proof. 
Thus, if the two vector fields are $\tau$ -related, the symmetry of $\Omega[\theta]=0$ is related to a symmetry of $\Sigma[\Phi, U]=0$, and so it induces an infinitesimal deformation of $\Phi$, and hence an integrated form of the surface is given by the expression
\be F=\Phi^{-1}pr \omega_R\Phi.\ee

In conclusion, we have shown that, based on the invariance criterion for generalized symmetries, a necessary and sufficient condition for existence of a Fokas-Gel'fand immersion function of a 2D soliton surface in a Lie algebra $\mathfrak{g}$ has been established. This condition is exactly that $pr\omega_R(\La\de{\cro{\theta},\la})=0$, whenever $\La\de{\cro{\theta},\la}=0$. We have given a geometric interpretation of this requirement in terms of the tangent vectors to the varieties defined by taking the quotient of the jet space with respect to the conditions $\Omega[\theta]=0$ and  $\La\de{\cro{\theta},\la}=0$, respectively. Furthermore, we have shown that it is sufficient to assume that  $\omega_R$ is $\tau$-related to $\eta_Q,$ which is a generalized symmetry of $\Sigma[\Phi, U]=0$, in order to show that the 2D-soliton surface can be integrated explicitly and the obtained expression \eref{eq19*} is consistent with tangent vectors \eref{tan}.  However, it is sometimes the case that a vector field $\omega_R$ is a symmetry of the ZCC $\Omega\cro{\theta}=0$ but not a symmetry of the LSP $\La\de{\cro{\theta},\la}=0$ since the action of $\prun{\omega_R}$ on the LSP does not vanish for all solutions $\Phi([\theta], \lambda)$ of the LSP. In this case, as long as the tangent vectors \eref{tan} are linearly independent, there  exists a $\mathfrak{g}$-valued immersion function $F$ for the 2D surface. However, the integrated form of the surface is not in the form given in \eref{eq19*}, i.e.
\begin{equation}
F\neq\Phi^{-1}\prun \omega_R\Phi,
\end{equation}
and consequently, not in the Fokas-Gel'fand form (\ref{f})
\begin{equation}
F\neq\Phi^{-1}{D\Phi\over D\theta^k}R^k.
\end{equation}

\noindent This phenomenon has been discussed in several examples (\cite{grundland2011soliton} - \cite{grundland2012soliton})   of either ordinary differential equations (ODEs) or partial differential equations (PDEs) concerning three possible forms of LSPs and their compatibility conditions.

\section{The elliptic equations and soliton surfaces}
We now present some examples which illustrate the theoretical considerations presented in Section 3. Consider a second-order autonomous ODE for a function $\theta$ in the independent variable $x$
\begin{equation}
\Omega[\theta]=2\theta_{xx} - f'(\theta)=0,\qquad f'(\theta)=\frac{d}{d\theta}f(\theta),\label{eq27*}
\end{equation}
where $f$ is some differentiable function of $\theta$. Equation \eref{eq27*} admits a first integral
\begin{equation}
\theta_x-\epsilon\sqrt{f(\theta)}=0,\qquad\epsilon^2=1,\label{eq28*}
\end{equation}
and the solution of \eref{eq28*} satisfies
\begin{equation}
\epsilon \int_{\theta_0}^\theta f(s)^{-1/2}ds=x-x_0.
\end{equation}
In the case when $(f(\theta))^{-1/2}=W(\theta,P_n(\theta)^{1/2})$, where $W$ is a rational function of its arguments and $P_n(\theta)$ is a polynomial of degree 3 or 4, the function $\theta$ which satisfies \eref{eq27*} is the inverse of an elliptic function \cite{byrd}. In particular, this is the case when $f(\theta)$ itself is a polynomial in $\theta$ of degree 3 or 4. The ODE \eref{eq27*} admits a matrix LSP for the function $\Phi$ in the Lie group $SL(2,\R)$ \cite{grundland2012surfaces}
\begin{eqnarray}
\label{eq29*}
&& D_x\Phi([\theta],y,\lambda)-U^1([\theta],\lambda)\Phi([\theta],y,\lambda) = 0 ,\\
&& D_y\Phi([\theta],y,\lambda)-U^2([\theta],\lambda)\Phi([\theta],y,\lambda) = 0, \nonumber 
\end{eqnarray} 
where the total derivatives \eref{total} in the direction of $x$ and $y$ take the form
\begin{equation}
D_x=\partial_x+\theta_x\frac{\partial}{\partial \theta}+\theta_{xx}\frac{\partial}{\partial \theta_x}+...,\qquad D_y=\frac{\partial}{\partial y}.
\end{equation}
and $\lambda\in\mathbb{R}$ is the spectral parameter. Note that we have introduced an auxiliary variable $y$ in the LSP \eref{eq29*} in such a way that the function $\Phi$ depends on $y$ but $U^1$ and $U^2$ do not (i.e. $D_yU^\alpha([\theta],\lambda)=0,$ $\alpha=1,2$). Consequently, the ZCC $\Delta[\theta]=0$ is equivalent to the Lax equation
\begin{equation}
D_yU^1-D_xU^2+[U^1,U^2]=D_xU^2+[U^2,U^1]=0\label{zcc}.
\end{equation}
The $sl(2,\R)$-valued matrix functions $U^\alpha$ satisfying \eref{zcc} are given \cite{grundland2012surfaces} by
\begin{equation}
U^1=\frac{1}{2}\left[\ba{rr}
0 & \frac{f'(\theta)}{\theta + \lambda}-\frac{f(\theta)-f(-\lambda)}{(\theta+\lambda)^2} \\
1 & 0
\ea\right],\quad U^2=\left[\ba{rr}
\theta_x & -\frac{f(\theta)-f(-\lambda)}{\theta+\lambda} \\
\theta +\lambda & -\theta_x
\ea\right],\label{eq32*}
\end{equation}
whenever $f(\cdot)$ is a polynomial in its argument.

Let us consider a particular solution of the LSP \eref{eq29*} with matrices given by (\ref{eq32*}), which can be expressed as follows \cite{grundland2012surfaces}
\begin{eqnarray}
\label{function}
\Phi =\left[\ba{rr}
2 D_x\psi_{+} &  2 D_x\psi_{-} \\
\psi_+ & \psi_- \ea\right] ,\\ 
\psi_{\pm} = \frac{\sqrt{ \theta + \lambda}}{\sqrt{2 a}} \exp\left(\pm a (y + \int_{\theta_0}^\theta \frac{ds}{2 \epsilon (s+ \lambda) \sqrt{f(s)}} )\right) ,\quad a= \sqrt{f(-\lambda)},
\end{eqnarray}
where the functions $\psi_{\pm}$ satisfy, in accordance with \eref{eq29*}, a system of differential equations

\begin{eqnarray}
D^2_x \psi_\pm = \frac{1}{4} \left( \frac{f'}{\theta - \lambda} - \frac{f - a^2}{(\theta + \lambda)^2} \right) \psi_\pm ,\label{bb}\\
 8a (D_x \psi_+) (D_x \psi_-) = \frac{f - a^2}{ \theta + \lambda} .\label{tt}
\end{eqnarray}
Equations (\ref{bb}) and (\ref{tt}) in terms of the function $\theta$ turn respectively into
\begin{eqnarray}\label{sistema1}
\Lambda([\theta],\lambda)=\left\lbrace\begin{array}{l}\frac{\theta_{xx}}{\theta + \lambda} - \frac{\theta_x^2}{2 (\theta + \lambda)^2 } -\frac{f'}{2(\theta + \lambda)} + \frac{f - a^2}{2(\theta + \lambda)^2}\\
+ a \left( \frac{\theta_{xx}}{(\theta + \lambda) \epsilon \sqrt{f}} - \frac{\theta_x^2 f'}{2 \epsilon (\theta + \lambda) f^{\frac{3}{2}}} \right) + \frac{a^2 \theta_x^2}{2 f (\theta + \lambda)^2 }=0  , \\ 
\theta_x^2  - a^2 \frac{\theta_x^2}{f}- f + a^2=0 . \end{array}\right .
\end{eqnarray}
Since the above equations have to hold for all values of $a$, it is straightforward to obtain from \eref{sistema1} the following set of differential equations for the LSP
\begin{equation}
\Lambda ([\theta],\lambda)=
\left\{\ba{r}
\theta_{xx} - \frac{1}{2} f'(\theta)=0 ,\\
\theta_x^2 - f(\theta) =0 .
\ea\right.
\end{equation}
This implies that a symmetry of $\Omega[\theta]$ is a common symmetry of the LSP if and only if it is also a symmetry of the differential equation $\theta_x^2 = f(\theta)$.   
The generalized symmetries of the ODE \eref{eq27*} are determined by the conditions \eref{deter} and \eref{Aalpha} leading to the following restriction on the function $R[\theta]$
\begin{equation}
2D_x^2R-f''(\theta)R=0,\label{eq35*}
\end{equation}
whenever $2\theta_{xx}-f'(\theta)=0$ holds. Hence, two particular solutions of the determining equation \eref{eq35*} are given by
\begin{eqnarray}
R^1 &=& \theta_x, \\
R^2 &=& \theta_x\int_{\theta_0}^\theta (f(s))^{-3/2}ds.
\end{eqnarray}
It is easy to verify that $R^1$ is a common symmetry of both the LSP and the ZCC

\begin{eqnarray}
pr \omega_{R^1} (\theta_{xx} - \frac{\theta_x^2}{2} ) = 0 , \quad \mbox{whenever} \quad \theta_{xx} - \frac{\theta_x^2}{2}=0, \\
 pr \omega_{R^1} (\theta_x^2 - f) =0 , \quad \mbox{whenever} \quad \theta_x^2 - f =0 ,
\end{eqnarray}

\noindent while for $R^2$ we obtain the following

\begin{eqnarray}
pr \omega_{R^2} (\theta_{xx} - \frac{\theta_x^2}{2} ) = 0  , \quad \mbox{whenever} \quad \theta_{xx} - \frac{\theta_x^2}{2}=0 , \\ 
pr \omega_{R^2} (\theta_x^2 - f)   \neq 0  , \quad \mbox{whenever} \quad \theta_x^2 - f =0 .
\end{eqnarray}
Hence, in this case, the integrated form of the surface associated with the symmetry $R^2$ is not of the form \eref{eq19*}, since the vector field $\omega_{R^2}$ is not a common symmetry of the initial system \eref{eq27*} and its LSP \eref{eq29*}.

\section{The $CP^{N-1}$ sigma model and soliton surfaces}

Let us now discuss the existence of soliton surfaces associated with the $CP^{N-1}$ sigma model which are obtained from the Fokas-Gel'fand formula for the immersion of 2D surfaces in a Lie algebra. For a complete review of the $CP^{N-1}$ sigma model, see \cite{GoldGrund2010}-\cite{mik}. In this section, we focus on the generalized symmetries of the integrable $CP^{N-1}$ equation and their relation to symmetries of the corresponding LSP. 

The integrable system $\Omega[\theta]=0$ for the $CP^{N-1}$ model is often given in terms of a rank-one Hermitian projector $P$ (satisfying $P^2=P, \, tr(P)=1, \, P^\dagger=P$) with Euler-Lagrange (E-L) equations
\be\label{EL} [\partial_+ \partial_- P, P]=\varnothing, \qquad \partial_\pm=\frac12\left(\partial_1\pm i\partial_2\right), \ee
where $\partial_1=\partial_{\xi^1}$, $\partial_2=\partial_{\xi^2}$. We assume that the model is defined on the Riemann sphere $S^2=\mathbb{C}\cup\lbrace\infty\rbrace$ and that its action functional is finite. It is well known that there exist raising and lowering operators for the solutions of the $CP^{N-1}$ sigma model and that any solution can be expressed as a raising operator acting on a holomorphic solution \cite{GoldGrund2010}-\cite{postgrundland}. The raising and lowering operators, in terms of the projectors, are given by 
\be \Pi_\pm(P)=\left\{\ba{lr} \frac{(\partial_\pm P)P (\partial_mp P)}{tr((\partial_\pm P)P (\partial_mp P))}& \mbox{ for }(\partial_\pm P)P \ne \varnothing,\\
 \varnothing & \mbox{ for }(\partial_\pm P)P = \varnothing\ea\right. .\ee
The raising and lowering operators constitute mappings between solutions of  (\ref{EL}). Successive applications of these operators give a set of orthogonal projectors and will eventually annihilate the operator, i.e. $\Pi^{k+1}_-(P)=\varnothing$ for some $k$. In this case, the projector $\Pi_-^k(P)$ is a holomorphic projector in the sense that it projects onto an equivalence class in $CP^{N-1}$ with a holomorphic representation \cite{postgrundland}. The operator $\Pi_-^k(P)$ is then said to be a k-th mixed solution. 
Finally, we complete our brief introduction to the $CP^{N-1}$ model by mentioning that the LSP is given \cite{mik}, \cite{zakharov} by 
\be U^{\pm}=\frac{2}{1\pm\lambda}[\partial_\pm P,P],\quad  \partial_\pm \Phi=U^{\pm}\Phi, \label{CPNLSPP}\ee
with solution
\be \Phi=I_N+\frac{4\lambda}{(1-\lambda)^2}\sum_{j=0}^{N}\Pi_-^{j}P-\frac{2}{1-\lambda}P\in SU(N), \qquad \lambda=it,\quad t\in\mathbb{R}\label{phi5},\ee
where $I_N$ is the $N\times N$ identity matrix. The surfaces associated with this integrable system have been investigated in several papers, including \cite{grundlandyurdusen} where it was proven that, for this system,  the Sym-Tafel formula for immersion is equivalent to the generalized Weierstrass formula for immersion. Surfaces associated with the symmetries of this model have also been  investigated \cite{grundland2011soliton}. In particular, it was shown that conformal symmetries of the integrable equation are symmetries of the LSP for the case of the sigma model defined above but not for traveling wave solutions of the sigma model defined on Minkowski space. It is the former case which we will focus on for the remainder of this section as an example of the theoretical considerations in Section \ref{theory}. 
 
For clarity, we express the model in a form consistent with the previous sections. Instead of using the projector $P$, we use an element of the Lie algebra $su(N)$ defined by 
\be \theta\equiv i(P-\frac{1}{N}I_N) \in su(N),\ee
with the algebraic restriction
\be \theta\cdot \theta=-i\frac{(2-N)}{N}\theta+\frac{(1-N)}{N^2}I_N, \label{algrestricttheta}\ee
corresponding to the requirement that $P^2=P$. The E-L equations become a set of differential equations for the functions $\theta^j$ given by 
\be \Omega^j[\theta]=\left[[(\partial_1^1+\partial_2^2)\theta, \theta]\right]^j=0,\ee
where $[\cdot]^j$ denotes the coefficient of the j-th basis element $e_j$ for the Lie algebra $su(N)$. 
The potential matrices $U^{\alpha}$ expressed in terms of $\theta$ are given by 
\begin{eqnarray} U^1=\frac{-2}{1-\lambda^2}\left([\partial_1\theta,\theta]-i\lambda[\partial_2\theta,\theta]\right)\in su(N)\\
U^2=\frac{-2}{1-\lambda^2}\left(i\lambda[\partial_1\theta,\theta]+[\partial_2\theta,\theta]\right)\in su(N),\end{eqnarray}
with the restriction that $\lambda$ be purely imaginary. 
Expressing the function $\Phi$ in terms of~$\theta$
\be \fl \Phi([\theta],\lambda)=I_N+\frac{4\lambda}{(1-\lambda)^2}\sum_{j=0}^{N}\Pi_-^{j}\left(\frac1NI_N-i\theta\right)-\frac{2}{1-\lambda}\left(\frac1NI_N-i\theta\right)\in SU(N)\label{Phitheta}\ee
reduces the LSP (\ref{LSP}) to a set of algebraic conditions on the jet space $[\theta]$
\be \Lambda^{\al j}([\theta],\lambda)=[(D_{\al} \Phi)\Phi^{-1}-U^{\al} ]^j=0,\qquad \al=1,2.\ee
In general these equations are very complicated due to the repeated application of the lowering operator. Depending on the dimension of the model, the equations may depend on many higher derivatives of the $\theta^j$, for example $\partial_{1}^2\partial_{2}\theta^{j}$  in the case of $N=3$. Thus, we see that in general these algebraic conditions on the jet space are not the same as $\Omega[\theta]=0$. Hence, even though the solutions of the differential equations $\Omega[\theta]=0$ and $\Lambda([\theta],\lambda)=0$ are the same, the algebraic constraints on the jet space are different, so it is possible that they have different symmetries. 

To give an explicit example, we consider the simplest case of the $\mathbb{C}P^{1}$ model. In this case, $\theta\in su(2).$ The algebraic requirement (\ref{algrestricttheta}) reduces the number of unknowns to two real functions $\theta^1$ and $\theta^2$, so that 
\be \theta=\left[\ba{cc} \sqrt{\theta^1 \theta^1+\theta^2\theta^2-\frac14}& \theta^1+i\theta^2\\
-\theta^1+i\theta^2 & -\sqrt{\theta^1 \theta^1+\theta^2\theta^2-\frac14}.\ea\right].\ee
These real functions can be expressed in terms of one unknown complex function $w$
\be \theta^1=\frac{w+\overline{w}}{2(1+w\overline{w})}, \qquad \theta^2= \frac{w-\overline{w}}{2i(1+w\overline{w})}.\ee
In this case, the E-L equation is given by 
\be \Omega[w]=\partial_1^2w+\partial_2^2w+\frac{1}{2(1+w\overline{w})}\overline{w}((\partial_1w)^2+(\partial_2w)^2)=0 \label{Omega1}.\ee
By directly substituting the form of the function $\Phi$ given by (\ref{Phitheta}) into (\ref{LSP}), we obtain the LSP as a function on the jet space $[\theta]$ 
\be \Lambda([w], \lambda)= \partial w\bar{\partial}w=0  . \label{Lambda1}\ee
That is, $w$ is either holomorphic or anti-holomorphic. On the one hand, we know from the general theory of solutions of $CP^{N-1}$ sigma models that if the functions to be considered are those defined on the Riemann sphere then (\ref{Omega1}) and (\ref{Lambda1}) have the same set of solutions.  On the other hand, these two algebraic conditions on jet space are not the same and so it is clear that the algebraic varieties associated with each of these equations, the spaces ${\cal O}$ and ${\cal L}$, are not the same. 

However, in this case  both equations admit conformal symmetries. That is, the vector fields
\be w_{C_1}=\partial_{\xi^1}w\frac{\partial}{\partial w} +\partial_{\xi^1}\overline{w}\frac{\partial}{\partial \overline{w}}, \qquad w_{C_2}= \partial_{\xi^2}w\frac{\partial}{\partial w} +\partial_{\xi^2}\overline{w}\frac{\partial}{\partial \overline{w}} \ee 
are symmetries of both $\Omega[w]=0$ and $\Lambda([w], \lambda)=0$. For simplicity of presentation, we have omitted the arbitrary function in the conformal symmetry.   In this case, the LSP also admits a conformal symmetry so that the vector fields $w_{C_1}$ and $w_{C_2}$ are $\tau$-related to the following vector fields, which are conformal symmetries of the LSP, 
\be \hspace{-1cm}\eta_{C_1}=\partial_{\xi^1}\Phi^j\frac{\partial}{\partial \Phi_j}+\partial_{\xi^1}U^{\al,j}\frac{\partial}{\partial U^{\al,j}},\qquad \eta_{C_2}=\partial_{\xi^2}\Phi^j\frac{\partial}{\partial \Phi_j}+\partial_{\xi^2}U^{\al ,j}\frac{\partial}{\partial U^{\al ,j}}.\ee
It is easy to directly verify the conditions (\ref{tau}) for which $\eta_{C_1}$ and $ \eta_{C_2}$ are $\tau$-related .

Here we have discussed the case $N=2$ in order to give an explicit example of the condition $\Lambda([\theta],\lambda)=0$ and to show the difference between the two varieties ${\cal O}$ and ${\cal L}$. The fact that, in general, a conformal symmetry of the $CP^{N-1}$ model is also a symmetry of the LSP was shown originally in  \cite{grundland2011soliton}. However, in this paper, we emphasize the geometric content of the concept of common symmetries and  the evaluation map as giving a relation between the vector fields.

\section{Concluding remarks}

In this paper we have shown how to determine common symmetries of two-dimensional integrable systems and their linear spectral problems in order to use the Fokas-Gel'fand formula for immersion of 2D-soliton surfaces. The existence of such common symmetries has been proven \cite{grundland2011soliton} to be a sufficient condition for the existence of the immersion function in the form postulated by the Fokas-Gel'fand formula \eref{f}. We have investigated the geometric consequences of this condition and rephrased it as a requirement for the corresponding vector fields to be group-related. Establishing an explicit expression for this relation has provided us with a tool for distinguishing between the cases in which the soliton surfaces can and cannot be integrated through the formula \eref{f}.

The problem of constructing integrable surfaces has remained one of the most actively researched subjects in mathematical physics in the last few decades. The usefulness of having an analytic description of surfaces generated by symmetries of the integrable system is well recognized. Geometric characteristics of these surfaces obviously facilitate the interpretation of the phenomena modeled by the investigated system. The advantage of this approach has increased recently with the development of computer techniques for the visualization of mathematical formulas. A visual image of a surface reflecting the behavior of solutions can be of interest, providing some clues about properties of this surface, otherwise hidden in some implicit mathematical expressions.

In this field the contribution of Fokas-Gel'fand et al. \cite{fokas2000aformula} is particularly valuable, providing a formula for the construction of integrable surfaces which is still the most general, workable and widely used. However, in practice one still encounters some examples \cite{grundland2011soliton,grundland2012surfaces} where it is not possible to construct integrable surfaces within this model. This raises the question of its applicability. In this paper we have addressed this question and formulated an easily verifiable condition which ensures that the Fokas-Gel'fand formula produces a desired result. We hope that this advance will assist future studies of 2D-surfaces associated with integrable models.

\section*{Acknowledgments}\label{acknowledg}

The authors thank Professor L. \v{S}nobl (Czech Technical University in Prague) for helpful and interesting discussions on the topic of this paper. This work was supported by a research grant from NSERC of Canada.

\end{document}